\RequirePackage{fix-cm}
\documentclass[smallextended]{svjour3} 
\usepackage{graphicx}
\usepackage{rotating}
\usepackage{amssymb}
\usepackage{mathptmx}
\usepackage{subfig}

\usepackage[sort,numbers]{natbib}
\newcommand{\aap}{A\&A}			
\smartqed
\makeatletter
\journalname{Journal of Low Temperature Physics}

\newcommand{\nat}{Nature}		
\newcommand{\apjs}{ApJS}		

\bibpunct{[}{]}{,}{s}{}{,}

\begin{document}

\newcommand{\hdblarrow}{H\makebox[0.9ex][l]{$\downdownarrows$}-}
\title{Calibration scheme for large Kinetic Inductance Detector Arrays based on Readout Frequency Response}

\author{L.Bisigello$^{1,2}$ \and S.J.C.Yates$^1$ \and V.Murugesan$^3$ \and J.J.A.Baselmans$^3$ \and A.M.Baryshev$^{1,2}$}

\institute{$^1$SRON Space Research of Netherland Groningen, 9747 AD, Groningen, The Netherlands\\$^2$ Kapteyn Astronomical Institute, University of Groningen, 9747 AD, Groningen, The Netherlands\\$^3$SRON Space Research of Netherland Utrecht, 3584 CA, Utrecht, The Netherlands\\
\email{l.bisigello@astro.rug.nl}}

\date{Submitted:30-09-2015, Accepted:22-01-2016}

\maketitle

\begin{abstract}

Microwave kinetic inductance detector (MKID) provides a way to build large ground based sub-mm instruments such as NIKA and A-MKID. For such instruments, therefore, it is important to understand and characterize the response to ensure good linearity and calibration over wide dynamic range.
We propose to use the MKID readout frequency response to determine the MKID responsivity to an input optical source power.  A signal can be measured in a KID as a change in the phase of the readout signal with respect to the KID resonant circle. 
Fundamentally, this phase change is due to a shift in the KID resonance frequency, in turn due to a radiation induced change in the quasiparticle number in the superconducting resonator. We show that shift in resonant frequency can be determined from the phase shift by using KID phase versus frequency dependence using a previously measured resonant frequency. Working in this calculated resonant frequency, we gain near linearity  and constant calibration to a constant optical signal applied in a wide range of operating points on the resonance and readout powers.\\
This calibration method has three particular advantages: first, it is fast enough to be used to calibrate large arrays, with pixel counts in the thousand of pixels; second, it is based on data that are already necessary to determine KID positions; third, it can be done without applying any optical source in front of the array.


\end{abstract}

\section{Introduction}
\begin{figure}
\begin{center}
\includegraphics[trim={0.5cm 0cm 1cm 1.5cm},clip,width=0.32\linewidth, height=0.26\linewidth ]{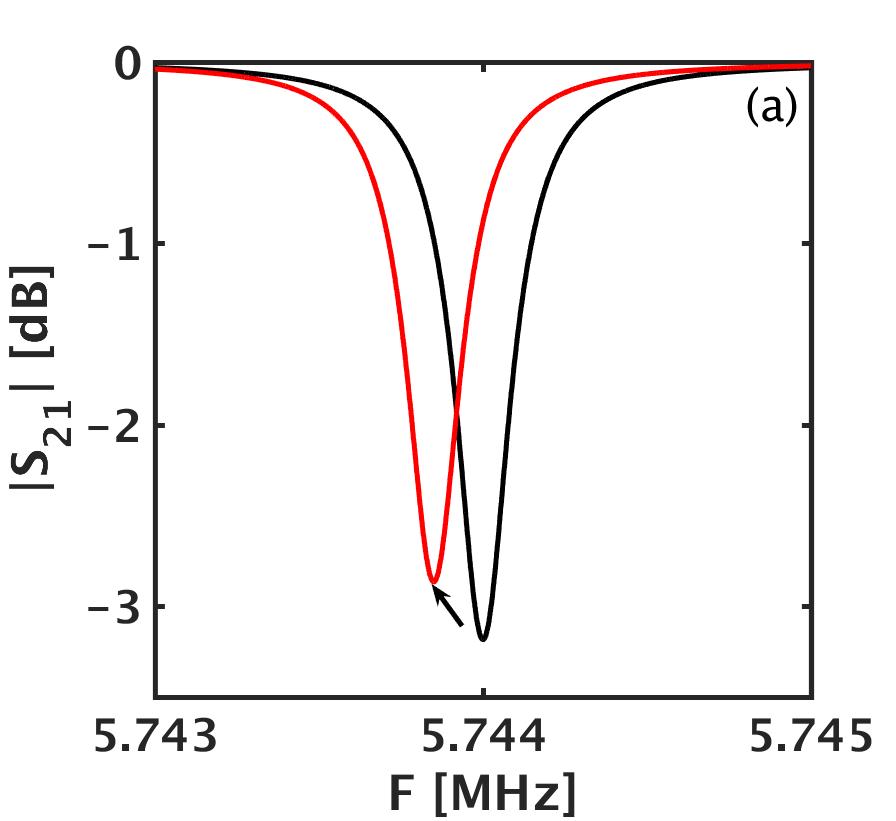}
\includegraphics[width=0.32\linewidth]{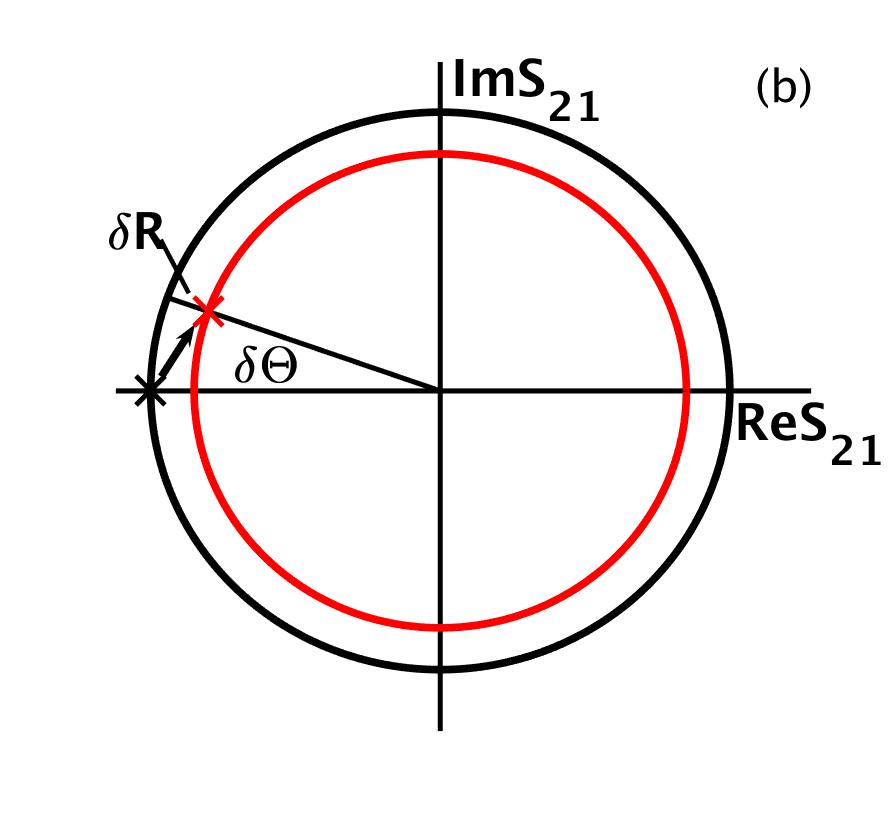}
\includegraphics[trim={0.5cm 1.5cm 2cm 2cm},clip,width=0.32\linewidth, height=0.28\linewidth ]{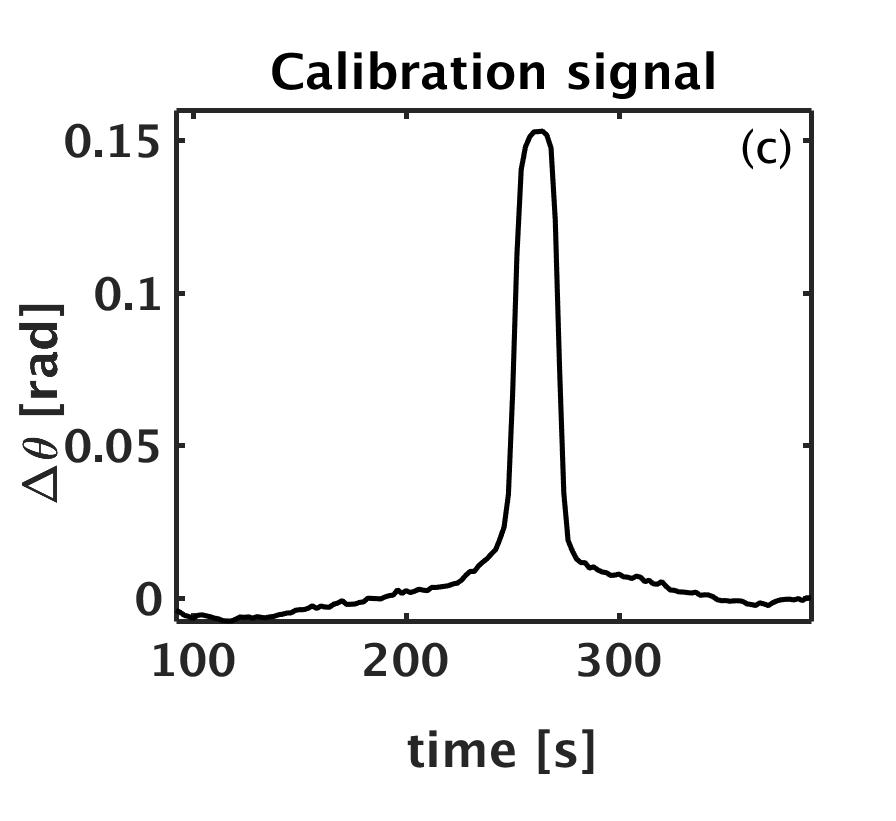}
\caption{(a) Transmission vs. readout frequency and (b) transmission in the complex plane. The black line is the transmission without any signal, while the red line is the transmission with a signal. (c) Readout phase versus time, as measured to calibrate with the calibrated optical signal. The feature is created at the moment when a gortex strip is between the optical source and the detector.}\label{manyfig}
\end{center}
\end{figure}

Microwave kinetic inductance detectors (MKID)\cite{Zmuidzinas2012,Baselmans2012,Day2003} are ideal for use in large ground-based sub-millimetre instruments, such as A-MKID\footnote{http://www3.mpifr-bonn.mpg.de/div/submmtech/bolometer/A-MKID/a-mkidmain.html} and NIKA\cite{monfardini2010,Monfardini2011}, because it is possible to read out simultaneously up to a thousand pixels with a single readout line\cite{vanRantwijk2015}.\\
When a photon is absorbed by a KID, it produces a change in the kinetic inductance. This change is visible both as a shift in the resonance frequency in the real plane (Fig. ~\ref{manyfig}a) and as a change in the transmission phase in the complex plane (Fig. ~\ref{manyfig}b). In particular, the change in the resonance frequency is proportional to the change in the kinetic inductance. Unfortunately, this shift in the resonance frequency is not directly measurable with a single fixed bias frequency, unless a modulation readout scheme is used\cite{Swenson2010,Calvo2013}. \\
There are different primary and secondary calibration methods, which allow for convert measured quantity, such as phase difference, to corresponding black body temperature difference and input signal. On sky, it is possible to use primary calibration sources, astronomical point sources, but it requires time since every pixel must measure this calibration source. An other way is to use a sky dip, where an elevation change of the telescope is used to calibrate the temperature scale. But this is not always possible, because it requires good magnetic shielding and knowledge of the sky transmission. In the lab, it is possible to calibrate an instrument with a polariser grid sweep, enabling going from 300K load to typically 77K by varying angle of polariser. Unfortunately, this grid can not be used to calibrate the entire telescope, because it is physically difficult to place this grid in front of the telescope, while also the calibration done in the lab may not be necessarily at loading condition used on sky. A secondary calibration source in the lab is a gortex sheet, which is slightly grey in band and is pre-calibrated with polariser grid. Calibration time depends on the calibration scheme, it requires more time to measure a source for every detector than to do a gortex sweep, which, therefore, is preferable.\\ Linearisation of the signal can be done by a polariser grid sweep, when available,  or sky dip but different working conditions, e.g. sky or elevation varying the load temperature, shift the resonance. A grid calibration can therefore become invalid or not the optimum operation point for best signal to noise. Therefore, having a calibration scheme base on the underlying operating principle, i.e. the KID resonant frequency change, enables more flexibility in particular to extrapolate between different operating points and the primary calibration. Also, such a scheme allows for linearisation in lab experiments where other schemes are not available. \\
This paper is organised as follows. In section 2, we describe the calibration model based on the KID resonance frequency change and we illustrate the experiment done to test this calibration method. In section 3, we explain and discuss our experimental results and, in section 4, we report our conclusions.

\section{Experimental details}

\begin{figure}
\begin{center}
\subfloat[]{\includegraphics[width=0.45\linewidth,height=0.35\linewidth]{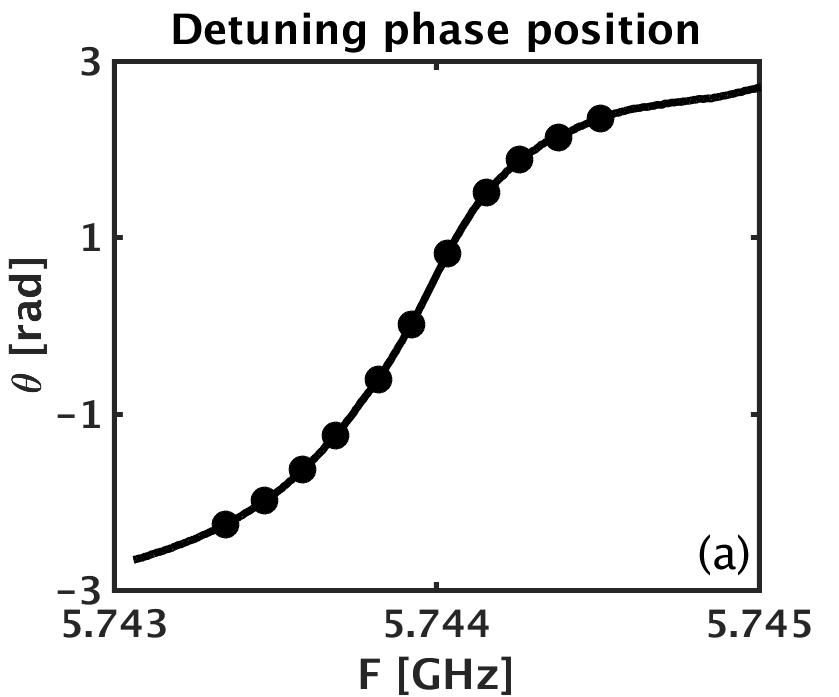} }
 \subfloat[]{\includegraphics[width=0.45\linewidth,height=0.36\linewidth]{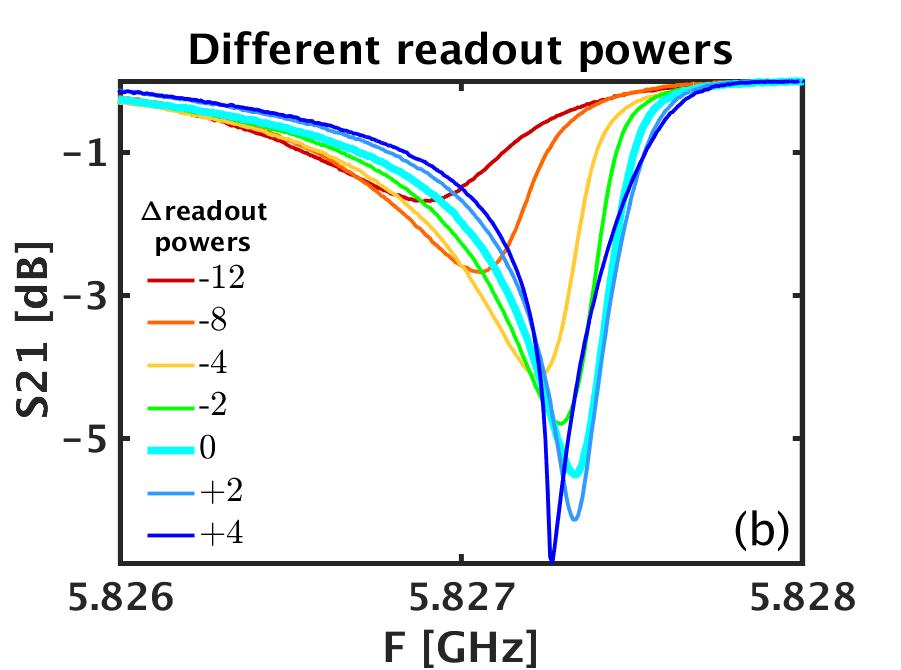} }
\end{center} 
\caption{Readout frequencies (a) and readout powers in dBm (b) used during the measurements. On the left the black line is the relation between phase and frequency, which could be approximated to a linear relation only within $\pm$1 rad.}
\label{detuning}
\end{figure}

We start from the assumption that the responsivity of a KID is:
\begin{equation}
\label{assumption}
	\frac{\partial f}{\partial T} = \frac{\partial f}{\partial \theta} \cdot \frac{\partial \theta}{\partial T}
\end{equation}
where f is readout frequency, T is the optical load temperature and $\theta$ is the phase transmission. $\partial f/\partial T$ is the responsivity,  $\partial f / \partial \theta$ can be obtained directly from the transmission by using the phase readout frequency relation(Fig.~\ref{detuning}(a)), while  $\partial \theta / \partial T$ can be calculated by using the calibration method based on calibrated optical signal. For very large changes in optical power, $\partial f/\partial T$ will have power dependence, but it can be assumed linear for these devices over a narrow range\cite{Calvo2013}. Under this condition, terms in the right part of the equation are inversely proportional and $\partial f / \partial \theta$  can be used to estimate  $\partial \theta / \partial T$. Therefore, we test under which condition this assumption holds experimentally. \\
We used a test system created for A-MKID that allow us to measure the readout phase and frequency of an input signal on a 880 pixel 350GHz A-MKID subarray.  We performed two sets of measurements, one to calibrate and the other to study the noise. We repeated all these measurements for different readout frequencies and different readout powers (Fig.  \ref{detuning}). The detuning positions are between $\pm$1 resonants bandwidths, which is the width of the resonance at half minimum dip depth. This range is wider than $\pm$1 rad that corresponds to quasi-linear regime in the phase-frequency relation, as it is visible in Fig.  \ref{detuning}(a), and that allows us to analyse linearisation far off resonance, i.e. large signal response. Here, we present the results of a representative KID. \\
First, we measured the signal of a liquid nitrogen background while moving a gortex strip in front of the array (Fig. \ref{manyfig}c) to partially obscured the signal. This obscuration had been previously calibrated to give a 21K signal difference on top of a liquid nitrogen background load by comparing to a polariser grid sweep between liquid nitrogen and 300K. In particular, we measured the output phase in two moments, when the optical source was directly observed and when the gortex strip was between the detector and the optical source. 
Since we previously calibrated the strip, we could calculate the derivative of the phase vs. the temperature (Fig. \ref{responsivity}(a)). We also calculated the derivative of the phase respect to the frequency (Fig. \ref{responsivity}(b)), in order to test the method based on readout frequency response. \\
Second, we measured the signal during an interval 40 seconds and we calculated the Power Spectral Density (PSD). From the PSD we evaluated the detector photon noise,  by subtracting from the measured noise the noise level above the KID roll-off, that corresponds to the amplifier contribution. Then, we divided the detector photon noise for the responsivity calculated with both methods in order to obtain the Noise Equivalent Temperature (NET) (Fig.  \ref{NET}).

\section{Experimental results and analysis}
\begin{figure}
\begin{center}
\includegraphics[trim={6cm 0cm 6cm 0cm},clip,width=1\linewidth,keepaspectratio]{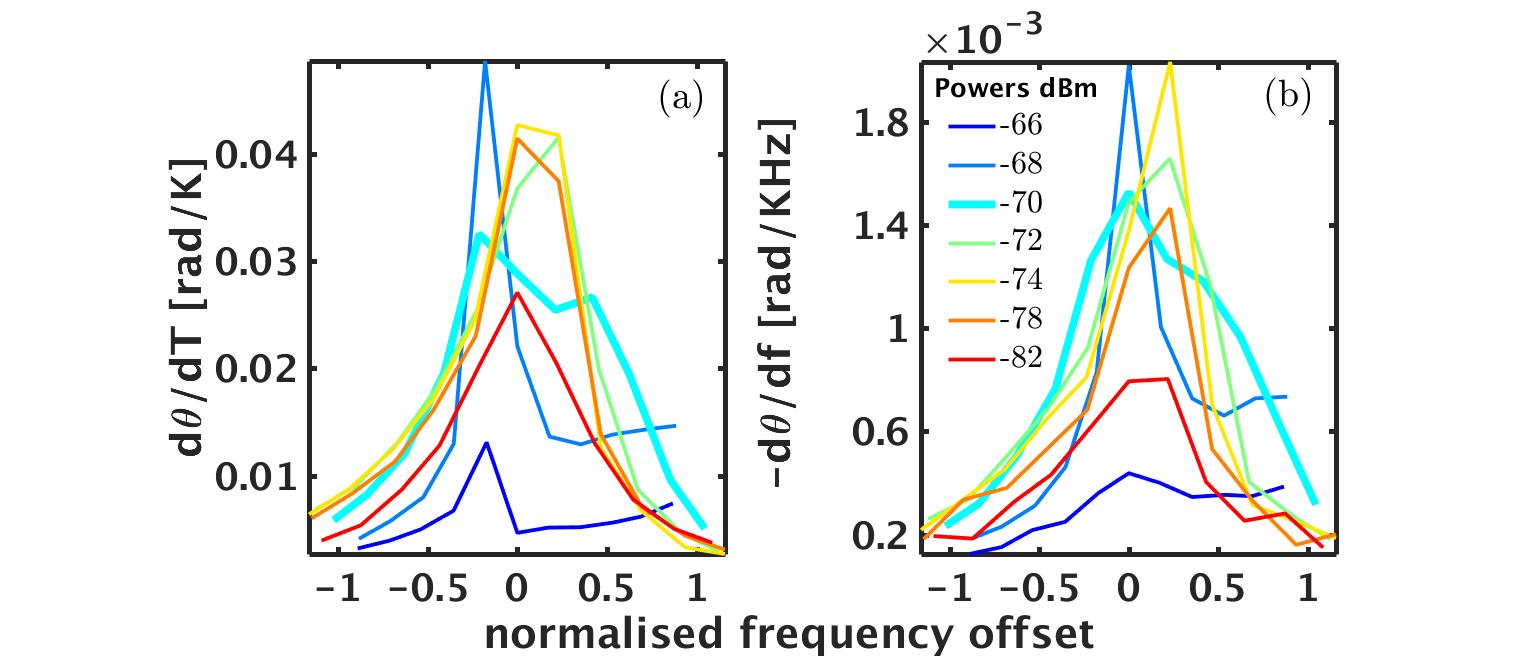} 
\end{center}
	\caption{Comparison between the measured responsivity $\frac{\partial \theta}{\partial T}$ (a) and the predicted responsivity $\frac{\partial \theta}{\partial f}$ (b). Colours correspond to different readout powers in dBm. Frequency offsets are normalised with respect to KID bandwidths.}
\label{responsivity}
\end{figure}
\begin{figure}
\begin{center}
\includegraphics[trim={2cm 0cm 1cm 3cm},clip,width=0.9\linewidth,keepaspectratio]{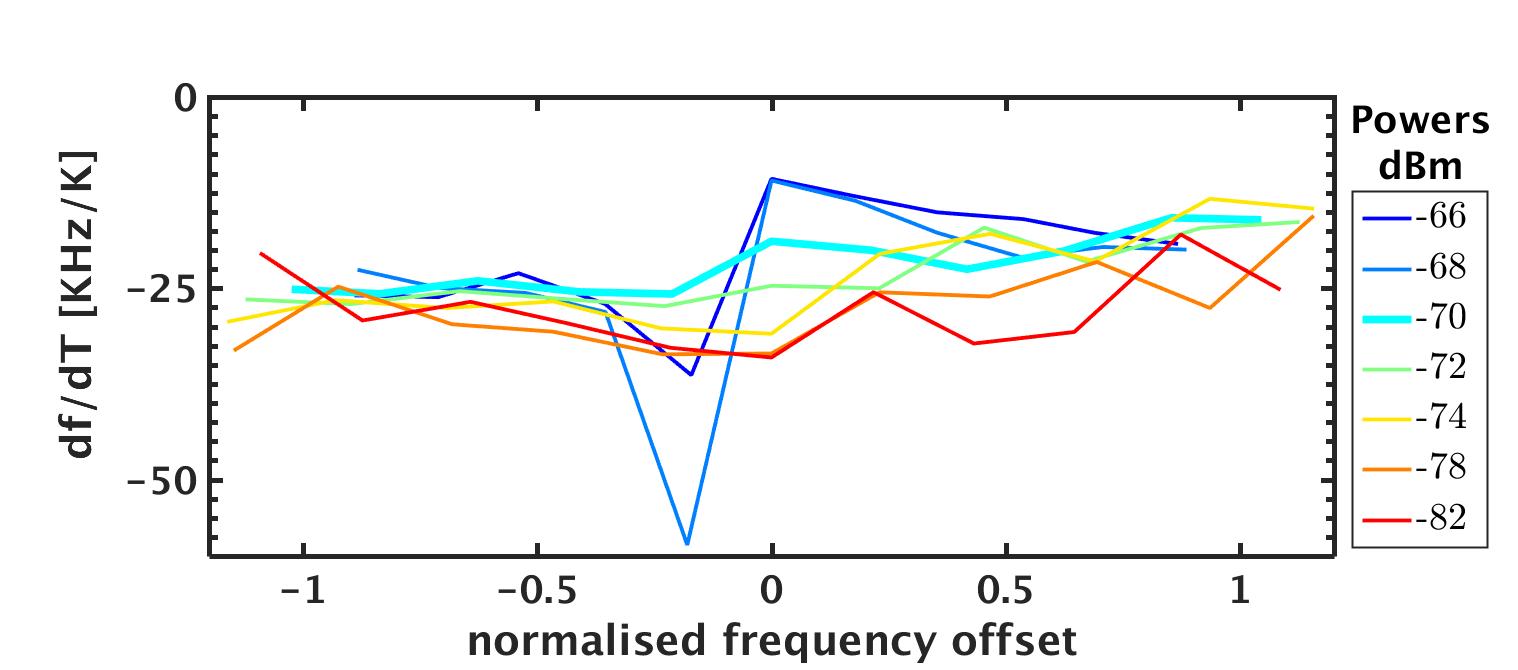} 
\end{center}
\caption{Responsivity $\frac{\partial f}{\partial T}$. It increases slightly with the readout powers and it is roughly constant with the readout frequency, but with some deviations for the highest readout powers. Colours correspond to different readout powers in dBm. Frequency offsets are normalised with respect to KID bandwidths.}
\label{ratio}
\end{figure}
\begin{figure}
\begin{center}
\includegraphics[trim={3cm 0cm 2cm 0.5cm},clip,width=0.9\linewidth,keepaspectratio]{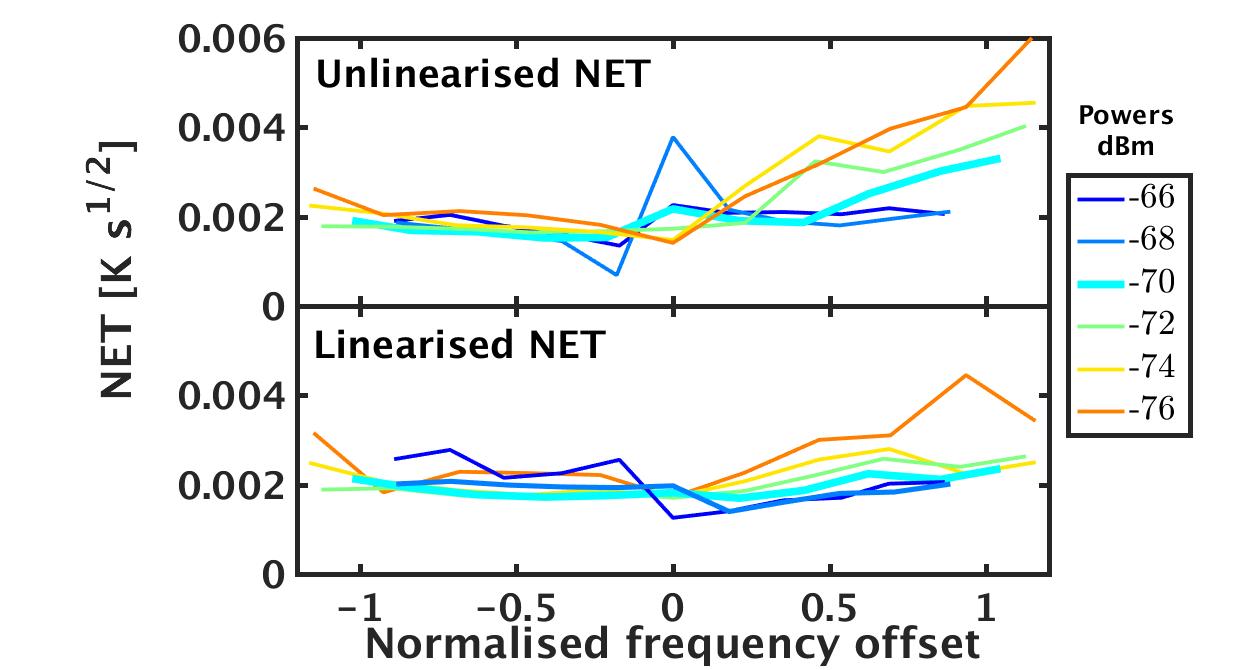} 
\end{center}
\caption{Comparison between the NET calculated by using a calibrated optical source and the NET predicted with the calibration method based on the readout frequency response. Colours correspond to different readout powers in dBm. Readout power gives some asymmetry also in the linearised NET, but the most symmetric ($\sim-$70 dBm) is linear to $\pm$10$\%$ over a very large range. Frequency offsets are normalised with respect to KID bandwidths.}
\label{NET}
\end{figure}

In figure \ref{responsivity}, it is shown the comparison between the responsivity measured with the method based on a calibrated optical source $\frac{\partial \theta}{\partial T}$ and the responsivity predicted with the method based on readout frequency response $\frac{\partial\theta}{\partial f}$. The phase responsivity  is clearly non-linear and it changes while varying the readout frequency and readout power. Therefore, before observing, it is important to know the combination of readout frequency and power that maximise the responsivity and, as a consequence, the performance of the instrument. The general shape of the optical responsivity is similar to that from the frequency dependence. The frequency responsivity, $\frac{\partial f }{\partial T}$ (Fig. \ref{ratio}), increases slightly with the readout powers and it is roughly constant with the readout frequency, but with some deviations for the highest and hence overdriven readout powers. \\
In figure \ref{NET}, it is shown the comparison between the NET obtained with the two methods. Readout power gives some asymmetry also in the linearised NET, but the most symmetric ($\sim-$70 dBm) is linear to $\pm$10$\%$ over a very large range.

\section{Conclusions}
We proposed and tested a calibration method based on MKID readout frequency response that could be use in large ground-based sub-millimetre instruments. This method as the advantage that is fast enough to be used in large arrays and it is based on data that are already used to measure KID positions.\\
We measured the responsivity based on the readout frequency response, $\frac{\partial\theta}{\partial f}$, and we confirmed that it is inversely proportional to the responsivity based on a calibrated optical source, $\frac{\partial \theta}{\partial T}$. Therefore, from a change in the phase response it is possible to predict a change in the resonance frequency and, therefore, a change in the kinetic inductance. In other words, this method can be used to calibrate MKIDs. Moreover, we calculated NET and we confirmed that this method allows for linearisation to $\pm$10$\%$ over a wide range in frequency and readout powers.

\begin{acknowledgements}
This project was supported by ERC starting grant ERC-2009-StG Grant 240602 TFPA and Netherlands Research School for Astronomy (NOVA).
\end{acknowledgements}


\end{document}